\documentclass[10pt,journal,compsoc]{IEEEtran}
\newif\ifpeerreview

\peerreviewfalse

\usepackage[nocompress]{cite}
\newcommand{\shortcite}{\cite}
\usepackage{amssymb}
\usepackage[switch]{lineno}
\usepackage[ruled]{algorithm2e} 

\SetAlFnt{\small}
\SetAlCapFnt{\small}
\SetAlCapNameFnt{\small}
\SetAlCapHSkip{0pt}
\IncMargin{-\parindent}

\usepackage{amsmath,amsfonts}
\usepackage{array}
\usepackage{textcomp}
\usepackage{stfloats}
\usepackage{url}
\usepackage{verbatim}
\usepackage{graphicx}
\usepackage{bbm}
\usepackage{xcolor}
\usepackage{subcaption}
\usepackage{multirow}
\usepackage{siunitx}
\usepackage{diagbox}
\usepackage{afterpage}
\usepackage[nolist,nohyperlinks]{acronym}
\usepackage{hyperref}
\usepackage{booktabs} 
\usepackage[section]{placeins}  
\usepackage{makecell}
\usepackage{pifont}
\usepackage{float}
\usepackage{etoolbox}

\newcommand{\ASM}{\text{ASM}}
\newcommand{\prop}{\text{Prop}}
\newcommand{\loss}{\mathcal{L}}
\newcommand{\org}{\mathbf{o}}
\newcommand{\dir}{\mathbf{d}}
\newcommand{\real}{\mathbb{R}}
\newcommand{\refr}{\text{refrac}}
\newcommand{\DOE}{\text{DOE}}
\newcommand{\comp}{\text{comp}}
\newcommand{\dO}{\emph{d\textbf{O}}}
\newcommand{\dB}{\mathrm{dB}}

\newcommand{\supp}[1]{{\textcolor{black}{#1}}}

\newcommand{\new}[1]{{\textcolor{black}{#1}}}
\newcommand{\final}[1]{{\textcolor{black}{#1}}}
\usepackage{titlesec}

\titlespacing*{\section}{0pt}{1.5ex}{1pt}
\titlespacing*{\subsection}{0pt}{1.2ex}{1pt}

\newcommand{\paperID}{31}

\title{Learned Off-aperture Encoding for \\Wide Field-of-view RGBD Imaging}

\author{%
Haoyu~Wei,%
~Xin~Liu,
~Yuhui~Liu,
~Qiang Fu,%
~Wolfgang~Heidrich,%
~Edmund~Y.~Lam,%
~and~Yifan Peng%
}

\begin{acronym}
  \acro{DNN}{deep neural networks}
  \acro{DOE}{diffractive optical element}
  \acro{SLM}{spatial light modulator}
  \acro{FoV}{field of view}
  \acro{WFoV}{wide field-of-view}
  \acro{AiF}{all-in-focus}
  \acro{PSF}{point spread functions}
  \acro{ASM}{angular spectrum method}
   \acro{FFT}{fast Fourier transform}
  \acro{LS-ASM}{least sampling ASM}
  \acro{E2E}{end-to-end}
  \acro{FT}{Fourier transform}
  \acro{DWA}{dynamic weight averaging}
  \acro{MSE}{mean squared error}
  \acro{EDoF}{extended depth-of-field}
  \acro{PSNR}{peak signal-to-noise ratio}
  \acro{EFL}{effective focal length}
  \acro{MAE}{mean absolute error}
  \acro{PMMA}{polymethyl methacrylate}
  \acro{OPL}{optical path length}
  \acro{CNC}{computer numerical control}
\end{acronym}

\begin{document}

\IEEEtitleabstractindextext{%

\vspace{-6pt}

\newcommand{\insertfig}{
\begin{center}\setcounter{figure}{0}
\includegraphics[width=0.95\textwidth]{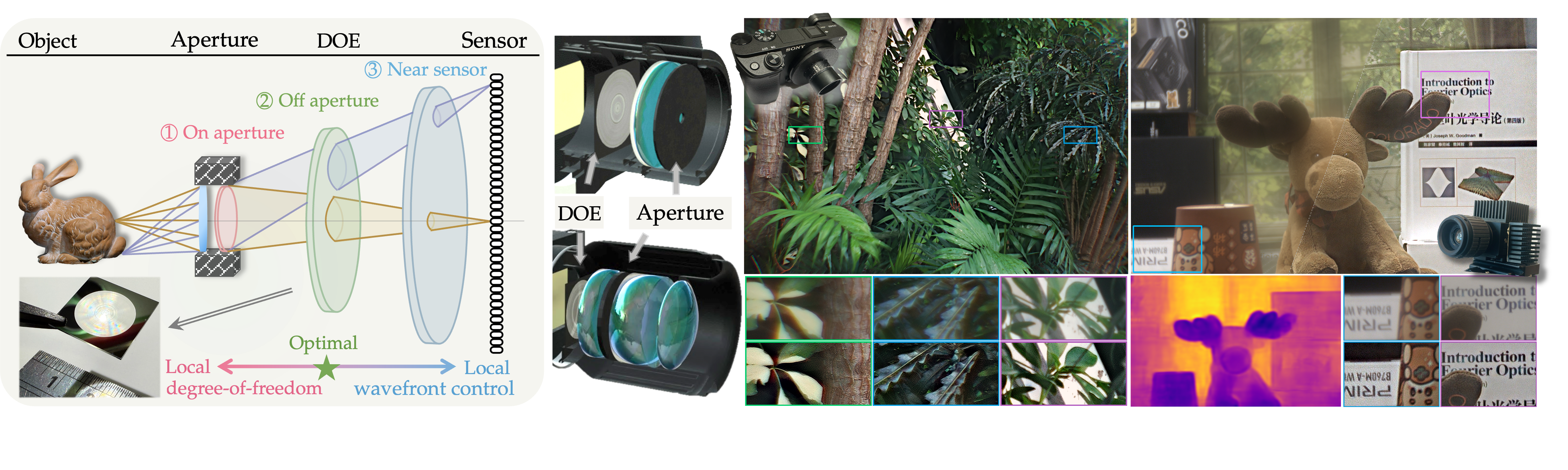}
\vspace{-35pt}
\captionof{figure}{
(Left) Depiction of three potential locations for integrating a \ac{DOE} for encoding purposes in an imaging system. The on-aperture DOE has many degrees of freedom to affect each ray bundle, but all of them are applied globally to the whole image plane. On the other hand, a DOE near the sensor provides localized control of the PSF, but with much fewer degrees of freedom for each ray bundle. The off-aperture DOE strikes an optimal balance between these two extremes.
(Center-left) Cross-sectional view of two custom-fabricated optical imaging systems. The DOEs and apertures are located at separate planes.
(Center-right) Resolved wide-\ac{FoV} results of App.1 (Sec.~5) compared to encoded measurements; (Right) Color and depth results of App.2 (Sec.~6).}
\label{fig:teaser}
\end{center}
\vspace{9pt}
}

\makeatletter

\insertfig

\begin{abstract}
End-to-end (E2E) designed imaging systems integrate coded optical designs with decoding algorithms to enhance imaging fidelity for diverse visual tasks.
However, existing E2E designs encounter significant
challenges in maintaining high image fidelity at wide fields of view,
due to high computational complexity, as well as difficulties in modeling
off-axis wave propagation while accounting for off-axis
aberrations. In particular, the common approach of placing the
encoding element into the aperture or pupil plane results in only a
global control of the wavefront. 
To overcome these limitations, this work explores an additional design
choice by positioning a \ac{DOE} \emph{off-aperture}, enabling a spatial
unmixing of the degrees of freedom and providing local control over
the wavefront over the image plane. 
Our approach further leverages hybrid refractive-diffractive optical systems by linking differentiable ray and wave optics modeling, thereby optimizing depth imaging quality and demonstrating system versatility.
Experimental results reveal that the off-aperture DOE enhances the imaging quality by over \qty{5}{\dB} in PSNR at a FoV of approximately \qty{45}{\degree} when paired with a simple thin lens, outperforming traditional on-aperture systems.
Furthermore, we successfully recover color and depth information at nearly \qty{28}{\degree} FoV using off-aperture DOE configurations with compound optics.
Physical prototypes for both applications validate the effectiveness
and versatility of the proposed method.
\end{abstract}


}

\maketitle

\IEEEdisplaynontitleabstractindextext

\IEEEraisesectionheading{
\section{Introduction}\label{sec:introduction}
}
\vspace{-5pt}
\IEEEPARstart{J}{\new{oint}}-designed optics, dubbed \emph{Deep Optics} in many research studies, have shown significant promise in a wide range of applications, including low-level vision tasks in photography~\cite{wu2019phasecam3d, shah2023tidy}, 3D vision and computational imaging~\cite{sitzmann2018end, jeon2019compact}, as well as holographic displays~\cite{gopakumar2024full}. Unlike traditional optical systems relying on heuristics~\cite{fischer2000optical}, deep optics can enhance the encoding capabilities of camera systems, allowing their measurements to be decoded through post-processing algorithms~\final{\cite{chakrabarti2016learning},}\cite{wetzstein2020inference, wang2025computational}. This encoding-decoding proficiency is jointly engineered to achieve optimal imaging objectives. Specifically, optical elements such as refractive lenses~\cite{sun2021end}, \ac{DOE}~\cite{sitzmann2018end,sun2020end,liu2022investigating}, and metalenses~\cite{tseng2021neural, gopakumar2024full} are incorporated into cameras to modulate the complex incoming wavefront. The parameters governing these optics, such as curvature, height map, or atom diameters, are updated in conjunction with differentiable algorithms employing backpropagation, typically neural networks~\cite{hu2024diffractive}.

In our work we are particularly interested in hybrid optics that
combine classical refractive lenses, which provide high diffraction
efficiency for focusing light, with diffractive optical elements (DOEs) for
optical encoding~\cite{yang2024end}. Research on this topic has
concentrated on optimizing the parameters of the encoding element,
while neglecting the question of where to place the DOE relative to the
aperture (or pupil plane). 
Conventional coded-aperture systems~\cite{zhou2011coded, levin2007image}, for simplicity, place the optical encoding on or near the aperture plane, creating \emph{global modulation for the entire wave field}.
For idealized unabberrated systems, this results in a shift-invariant PSF across the image plane. 
However, in the presence of off-axis aberrations, the global influence
of the DOE parameters on the whole image becomes \emph{an obstacle for
localized aberration correction}. 
To address these issues, we explore separating the encoding plane from the aperture plane, so that different regions of the DOE produce different encoding characteristics based on the incoming light directions.

It is worth noting that in most imaging scenarios where the sensor
size significantly surpasses that of the pupil, placing the \ac{DOE}
closer to the sensor increases the total degrees of freedom as well as
the locality of the wavefront control. Nonetheless, positioning the
DOE directly on the sensor is also not optimal, since the ability to
spatially redistribute light is diminished in this position
(Fig.~\ref{fig:teaser}).
In this work, we investigate this inherent trade-off between the locality of
control and the ability to redirect light.

The localized control over the wavefront shape is particularly relevant for wide-FoV imaging, where off-axis aberrations require a localized phase modulation for PSF shaping. 
In this work, we first show that \new{in compact imager settings,} pairing a simple thin lens with an off-aperture DOE enables the recovery of aberration-free images at $45^\circ$, outperforming the on-aperture DOE system by over $5~\dB$ in \ac{PSNR}.
Expanding on this achievement, we introduce more complex compound optics to tackle the challenging task of simultaneous color imaging and depth recovery at a wide-FoV of around $28^\circ$. 
This extension leverages the co-optimization of an off-aperture DOE with a Cooke triplet configuration~\cite{fischer2000optical}---comprising three refractive lenses positioned near the aperture plane, which partially corrects large-angle aberrations.
This setting utilizes the high diffraction efficiency of refractive lenses to focus light, while allowing the DOE to encode high-frequency information essential for depth estimation.
This hybrid refractive-diffractive system is modeled in two steps; the
differentiable ray-tracing engine \dO~\cite{wang2022differentiable} is
used for light propagation through the refractive lenses, followed by two-step off-axis wave propagation utilizing the \ac{LS-ASM}~\cite{wei2023modeling}. 
This hybrid design preserves the precise diffractive attributes of the optics, a feature absent in purely refraction-based systems.

As we showcase applications of multiple optimization objectives, we devise a lightweight multi-head neural network architecture that effectively extracts different information channels (e.g., color and depth) through separate decoding heads. Our design integrates a shared pre-trained feature extractor, with individual heads dedicated to specific tasks. This framework not only alleviates the training burden compared to conventional single-head networks~\cite{ikoma2021depth}, but also facilitates efficient information encoding.
In summary, our contributions are as follows:
\vspace{-5pt}
\begin{itemize}
\item We present a co-optimized off-aperture encoding framework tailored for wide-FoV computational imaging applications. Notably, we explore the critical impact of off-aperture \ac{DOE} placement. 
    
\item Our approach involves a differentiable refractive-diffractive hybrid imaging pipeline that integrates accurate off-axis wave propagation modeling, enabling the co-design of an off-aperture \ac{DOE} and a multi-head image-processing network. 
    
\item Through our simulations, we demonstrate high-fidelity lightweight lens imaging capability at \qty{45}{\degree} with a \ac{PSNR} gain over $5~\dB$ compared to on-aperture systems, and RGBD imaging capability at \qty{28}{\degree}. 

\item We develop two camera prototypes: one equipped with an off-the-shelf convex lens, while the other features a customized compound lens incorporating bespoke \ac{CNC}-turned aspherical lens geometry. Both prototypes utilize custom nanofabricated DOEs as encoding elements. Our systems are rigorously validated using a diverse range of indoor and outdoor scenarios.
\end{itemize}

\section{Related Work}
\label{sec:relatedwork}


\textbf{Coded-Aperture Computational Imaging.}
Conventional image system design optimizes optics and algorithms separately~\cite{cossairt2010spectral, zhou2011coded, zhang2021ten}, overlooking their potential synergistic interaction within a unified system.
Computational imaging addresses this limitation by jointly optimizing encoding optics and decoding algorithms.
Recent advances in AI, particularly deep neural networks, have facilitated the E2E design of computational imaging systems, co-optimizing hardware and software for specific tasks. 
These systems employ encoding masks to modulate incident light's amplitude, phase, and polarization, capturing rich scene information. Representative applications include hyperspectral imaging~\cite{monakhova2020spectral, shi2024learned}, superresolution~\cite{sun2020end}, extended depth-of-field~\cite{sitzmann2018end, seong2023e2e}, and depth estimation~\cite{wu2019phasecam3d,tan2021codedstereo}.

\new{Several off-aperture designs have been studied in other imagers in the past decade. A near-sensor amplitude mask was analytically designed for light-field encoding~\cite{veeraraghavan2007dappled}. Another work placed a DOE at an optimized position in a 4f system, using approximate off-axis wave simulations~\cite{ferdman2022diffractive}. Concurrently, a metalens positioned off the aperture stop was optimized to maximize the pupil diameter for telescopic imaging at \qty{20}{\degree} FoV, simulated with geometric optics~\cite{wang2025portable}. Unlike these approaches, our work aims to systematically explore and characterize this additional design space for wide-FoV compact imagers.}

\begin{figure*}[ht]
  \centering
  \includegraphics[width=\linewidth]{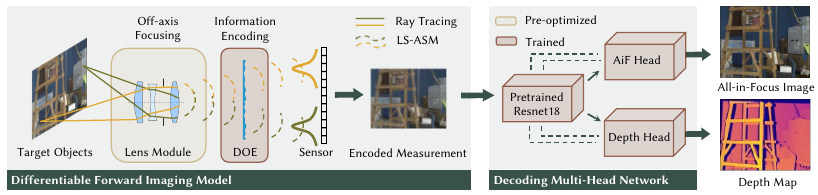}
  \vspace{-12pt}
  \caption{Imaging pipeline of our proposed system. We model the camera's light propagation using a combination of ray tracing and wave propagation. The compound lens module is pre-optimized and kept fixed during training, while the entire system is fully differentiable. The \ac{DOE} placed in-between the lens module and the sensor facilitates information encoding. A multi-head decoding network based on the ResNet architecture is incorporated to support optimization for multiple visual tasks.
  }
  \label{fig:main-pipeline}
  \vspace{-9pt}
\end{figure*}

\vspace{3pt}
\noindent \textbf{Wide Field-of-View Imaging.}
Wide-angle imaging settings are particularly challenging for optical systems with reduced complexity due to significant off-axis aberrations.
Notably, efforts to recover sharp color and depth information from defocused images~\cite{liu2022investigating, ikoma2021depth} often assume spatial shift-invariance, valid only within a limited \ac{FoV}.
Expanding the \ac{FoV} requires accurate modeling and mitigation of off-axis aberrations inherent in all imaging systems.
Peng et al.~\shortcite{peng2019learned} successfully demonstrated an imaging system that supports up to $53^\circ$ by first crafting the shift-variant \ac{PSF} of a thin-plate lens using ray-tracing. A separate post-processing network was designed to remove residual artifacts.
Nevertheless, commercial optical design software is often inefficient in modeling wave optics, particularly at large angles~\cite{matsushima2010shifted}. Moreover, task-driven designs require significant manual effort when using these tools, due to their incompatibility with E2E optimization.
Differentiable ray-tracing engines have also been used to design compound optics systems enabling end-to-end \ac{AiF} imaging over a $30^\circ$ \ac{FoV}~\cite{sun2021end,tseng2021differentiable,yang2024curriculum}\new{\cite{teh2024aperture}}.
Despite their effectiveness, compound optics tend to be bulky and have limited bandwidth to encode additional scene information.
\new{A comprehensive summary of representative encoded imaging literatures can be found in Supplementary Table.~\supp{1}.}

\vspace{3pt}
\noindent \textbf{Learning-based Image and Depth Recovery.}
Decoding both images and depth from a single encoded image is an ill-posed problem, typically addressed using deep neural networks.
Standard U-Net or attention U-Net~\cite{oktay2018attention} with RGB-D output channels have been utilized in coded-aperture systems~\cite{ikoma2021depth, shah2023tidy}. However, the disparate features and sparsity of color images and depth maps impose significant burdens on the layers preceding the output.
Anwar et al.~\shortcite{anwar2017depth} predicted depth and image sequentially, with the deconvolution performed using the intermediate depth predictions. This approach prioritizes depth estimation, as its quality is crucial for image deblurring.

More recently, a two-head network decouples the depth and image estimation tasks using parallel decoders sharing a feature encoder~\cite{nazir2023depth}. This allows for task-specific decoder designs but results in a larger model that can be difficult to train and susceptible to gradient vanishing.
Therefore, to mitigate memory demands, we adopt a pre-trained ResNet18~\cite{he2016deep} as the lightweight feature extractor with skip residual connections linking the encoder and decoders~\cite{godard2019digging}.

\vspace{3pt}
\noindent \textbf{Refractive-Diffractive Hybrid Optics.}
Combining simple thin lenses with diffractive lenses offers compact and efficient encoding for coded aperture applications~\cite{pinilla2022hybrid, evdokimova2022hybrid, ikoma2021depth}. However, refractive components are still modeled with simple analytical functions, such as the Gaussian lens formula~\cite{fischer2000optical}, excluding the use of compound optics and limiting the development of more advanced systems. This limitation is particularly critical for wide-FoV imaging, where simple lenses suffer severe off-axis aberrations.

The most recent studies explore a differentiable hybrid lens design and showcase the potential of aberration correction, depth estimation, and depth-of-field extension. 
Yang et al.~\shortcite{yang2024end} represented a significant
advancement in fully E2E optimizing both the optics and
image-processing algorithms. However, their learned \ac{DOE} also acts
as the aperture stop of the overall system, whereas each \ac{DOE}
feature {\em globally} affects the wavefront of the entire image, similar to conventional coded aperture methods.
Zhuge et al.~\shortcite{zhuge2024calibration} similarly optimized a Cooke triplet with a DOE but their DOE is, again, a near-aperture design. Furthermore, these works are limited to RGBD imaging with a FoV of under \qty{20}{\degree}.
In comparison, our work seeks to separate the aperture and
encoding planes of the optical system and to comprehensively investigate the impact of
\ac{DOE} positioning for the overall system performance, especially for
wide \ac{FoV} systems. Furthermore, we have experimentally
demonstrated not only wider \ac{FoV} imaging, but also extra depth
estimation capabilities, compared to recent counterparts.

\section{Shift-Variant Hybrid Imaging Model}

The optical part of our imaging system is composed of refractive lenses and a DOE, as illustrated on the left of Fig.~\ref{fig:main-pipeline}. We demonstrate the design paradigm with two configurations tailored for specific imaging tasks. 
Initially, as the foundational application, we employ a simple main lens and a DOE placed off-aperture to realize wide-FoV imaging. 
Expanding upon this, the second application aims at not only enabling wide-FoV imaging capability, but also exploiting depth estimation (Fig.~\ref{fig:main-pipeline} right), facilitated by a pre-optimized Cooke triplet and a co-designed DOE. 

To co-design these optical elements and image-processing neural networks, we establish an image formation model that simulates light propagation from the object to the image sensor. 
In this model, we first resort to geometric optics to obtain the response, i.e., the modified complex amplitude, of the refractive lenses (Sec.~\ref{sec:geo-optics-sim}). 
Specifically, we assume the propagation through the simple lens is achromatic and model it by wavelength-dependent quadratic phase, while that of the compound optics is derived by ray tracing. 
Upon obtaining the response, we employ wave optics theory to accurately calculate the \ac{PSF}s following \ac{DOE} modulation (Sec.~\ref{sec:off-aperture-modeling}). 
Finally, wide-FoV images are synthesized through spatial shift-variant convolution between the object and the \ac{PSF}s.

\subsection{Geometric Optics Simulation}
\label{sec:geo-optics-sim}

Complex lens elements are widely employed in imaging systems to enhance performance, particularly in mitigating off-axis aberrations. To accommodate such design scenarios, we simulate complex lenses with geometric optics.

The lenses can be optimized, or predefined and then simulated using a differentiable ray tracing engine \dO~\cite{wang2022differentiable} to obtain the complex field before wave optics is applied. Rays are traced from the front to the backplane of the lens module's bounding box, which is tangential to the exterior of the lens module and assumed to be fully refractive within its boundaries.
The backplane functions as the transition plane, linking refractive and diffractive optics.
The complex fields at this plane represent the transfer functions of the compound lens at various angles.

The ray tracing process involves tracing rays between consecutive surfaces of the lenses, the aperture, and the front and back planes.
At each point $\org_i\in\real^3$ on the surface $i$, where $i=0,\dots,S-2$ and $S$ is the total number of surfaces including the backplane, a ray $\{\org_i, \dir_i\}$ intersects the subsequent surface $i+1$ at $\org_{i+1} = \org_i+t_{i} \dir_i$. 
Here, the direction $\dir_i\in\real^3$ is a unit vector determined by the previous ray direction $\dir_{i-1}$ and the surface material, following Snell's law. 
Notably, for $i=0$, $\dir_{i-1}$ corresponds to the incident light direction. The ray marching distance $t_i\in\real^+$ is computed at surface intersections on $i$ and $i+1$ utilizing iterative root-finding techniques such as Newton's method.

The total phase shift of a ray is associated with the \ac{OPL} through the lenses, given by the sum of \ac{OPL}s $n_i t_i$ in each propagation medium, \final{where $n_i$ is the refractive index of the medium from surface $i$ to $i+1$.}
The rays reach scattered points on the transition plane, but subsequent processes such as wave propagation or sensing require measurements on a uniform grid. 
Thus, the points are interpolated onto a regular grid $(\xi, \eta)$ using Clough-Tocher interpolation~\cite{alfeld1984trivariate}, denoted as $g$. The phase $\phi_{\refr}$ at the wavelength $\lambda$ is then expressed as
\vspace{-4pt}
\begin{equation}
    \phi_{\refr}(\xi,\eta)=\frac{2\pi}{\lambda} g\left(\sum_{i=0}^{S-2}n_i t_i \Big| \xi, \eta\right).
\end{equation}
\vspace{-3pt}

Consistent with prior works~\cite{wang2022differentiable, yang2024curriculum}, we ignore energy decay in geometric light propagation and assume uniform energy for each ray. 
The intensity $I_{\refr}$ at the transition plane is obtained by bilinearly interpolating to adjacent grid points, accumulating ray energies. The transfer function is defined as $E_{\refr} = \sqrt{I_{\refr}}\exp\left(j\phi_{\refr}\right)$, where $j=\sqrt{-1}$.

\subsection{Off-Aperture Diffraction Modeling}
\label{sec:off-aperture-modeling}
Our two-step wave propagation process begins at the transition plane, traverses the off-aperture \ac{DOE}, and ultimately culminates at the sensor plane.
To simulate the highly off-axis wave propagation, we employ the well-established \ac{ASM}~\cite{goodman2017introduction}. The propagation of an input field $E_{\text{in}}(\xi,\eta)$ to a finite region on plane $(x,y)$ over a distance $z$ is expressed as:
\begin{equation}
    \ASM(E_{\text{in}} | \xi,\eta;x,y) = \mathcal{F}^{-1} \left\{\mathcal{F}\left\{E_{\text{in}}\right\} H(f_X,f_Y) \right\},
\end{equation}
where $H(f_X,f_Y)=\exp\left[jkz\sqrt{1-(\lambda f_X)^2 - (\lambda f_Y)^2}\right]$ is the frequency-domain transfer function, $k=2\pi/\lambda$ is the wave number, and $\mathcal{F}$ denotes the Fourier transform operator.

The off-aperture \ac{DOE} connects two wave propagation pathways at large angles, which can be computationally demanding. To address this, we employ the \ac{LS-ASM}~\cite{wei2023modeling} that introduces a compensation term to the phase of the input field, resulting in a quasi-on-axis field. This reduces the sampling requirements for simulating off-axis propagation.
For an input field at coordinates $(x,y)$, the compensation term is defined as $\phi_{\comp}(x,y)=-2\pi (f_{X_\text{c}} \xi+f_{Y_\text{c}} \eta)$, where $(f_{X_\text{c}}, f_{Y_\text{c}})$ represents the effective spectrum center.
The wave field arriving at the \ac{DOE} plane is then computed as:
\begin{equation}
    E_{\DOE}(\hat p,\hat q) = \ASM \big\{ E_{\text{in}}(\xi,\eta) \exp [j\phi_{\comp}(\xi,\eta)] \big\}.
\end{equation}

The learned \ac{DOE} globally modulates the incident field with $\phi_{\DOE}(p,q) = k (n_{\lambda_0}-1) h(p,q)$, where $n_{\lambda_0}$ is the refractive index of the substrate material at the nominal wavelength, and $h$ is the height map of the designed \ac{DOE}.
However, only a window at $(\hat p,\hat q)$ is calculated on the \ac{DOE} based on the chief ray's origin $\org_{S-1}$ and direction $\dir_{S-1}$ at $\lambda_0$ from the lens module to encompass the majority of diffracted energy. 
\final{Note that the DOE is also compensated.}
The complete wave propagation is thus formulated as:
\vspace{-4pt}
\begin{align}
\begin{split}
    \textbf{\prop} (E_{\text{in}}|\xi,\eta; &p,q; x,y) = \ASM\Big\{ E_{\DOE}(\hat p,\hat q) \times\\
    & \exp \left\{j[\phi_{\comp}(\hat p,\hat q) + \phi_{\DOE}(\hat p,\hat q)]\right\}\Big\}.
\end{split}
\end{align}

For \new{our case,} a point light source at $(x_0, y_0, z_0)$ \new{results in $E_{\text{in}}(\xi,\eta) = E_{\refr}\left(\xi,\eta\right)\exp\left(jkr\right)$, thereby} the \ac{PSF} is computed as the squared amplitude of the complex field arriving at the sensor plane within the specified window $(x,y)$:
\vspace{-4pt}
\begin{equation}
    K(x,y) = \left\lvert \textbf{\prop}\left\{E_{\refr}\left(\xi,\eta\right)\exp\left(jkr\right)\right\} \right\rvert^2,
\end{equation}
where $r = \sqrt{(x_0-\xi)^2 + (y_0-\eta)^2 + z_0^2}$.
Refer to Fig.~S1 in the supplementary material for details of the propagation modeling.

To simulate camera measurements, we convolve the PSFs with local image patches, assuming local shift-invariance due to the relatively smooth variations of PSFs across the \ac{FoV}.
We consider multiple depths $z_0=0,\dots,D-1$ and utilize the established occlusion-aware model~\cite{ikoma2021depth} to provide continuous and seamless representation of depth variations:
\vspace{-4pt}
\begin{equation}
     I = \sum_{d=0}^{D-1} \frac{1}{U_d}(K_d \ast I_d) \prod_{d'=d+1}^{D-1} (1-\tilde\alpha_{d'}) + \mathcal{N},
\end{equation}
where $\tilde \alpha_{d'}=(K_d \ast \alpha_d) / U_d$, $U_d = K_d \ast \sum_{d'=0}^d \alpha_{d'}$, $\ast$ denotes element-wise multiplication, \final{and $K_d$ is the sensor field from a light source at depth $z_0=d$}. The images are quantized into $D$ depth layers denoted as $I_d$, accompanied by a binary mask $\alpha_d$. Each depth layer is normalized by $U_d$, and Gaussian noise $\mathcal{N}$ is incorporated into the simulation. Sensor spectral response is also applied to image $I$ to address the sensor's sensitivity to different wavelengths.

\section{Multi-Head Image-Processing Network}

\subsection{Network Architecture}

The reconstruction network is tasked with extracting compressed information from sensor measurements and decoding it into both image and depth data (the right in Fig.~\ref{fig:main-pipeline}). However, color images and depth maps possess fundamentally distinct spatial characteristics: images are characterized by dense color features, while depth maps exhibit sparse features. 
Moreover, sharp image reconstruction promotes the \ac{DOE} to generate similar \ac{PSF}s across various depths, whereas depth estimation prefers distinct \ac{PSF} variations along the depth axis. This inherent difference necessitates extracting different types of information from the encoded features. Consequently, it is essential to employ separate decoding heads for each task, as illustrated in prior work~\cite{nazir2023depth}.

To streamline training and reduce computational overhead, we adopt a shared feature extractor for both decoding heads. Specifically, we leverage a lightweight pre-trained ResNet18~\cite{he2016deep}, similar to Monodepth2~\cite{godard2019digging}, and fine-tune it to better align with the encoded features in the sensor measurements. 
Each decoding head is comprised of four upsampling blocks, with each block integrating bilinear interpolation and residual convolution layers. Additionally, skip connections are applied between the ResNet encoder's feature maps and the corresponding layers of decoders, facilitating efficient gradient flow and enhanced feature decoding across all levels.

\begin{figure*}[t]
    \centering
    \includegraphics[width=0.99\linewidth]{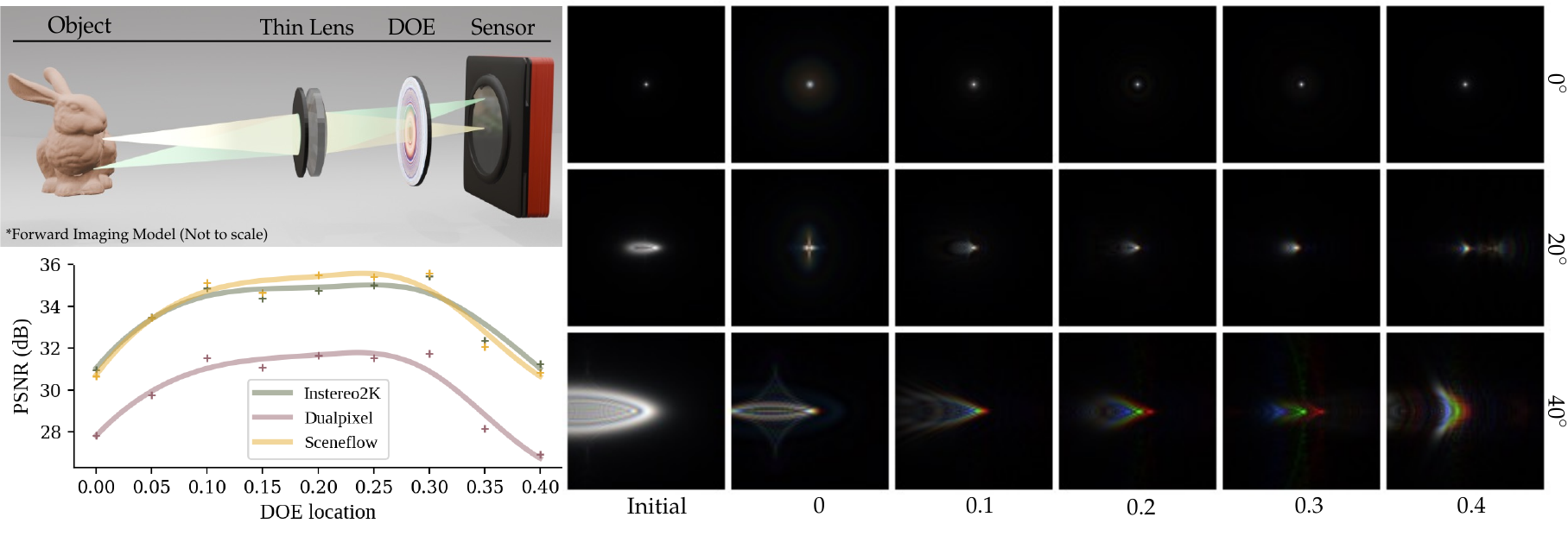}
    \vspace{-6pt}
    \caption{
    \final{(Top-left)} The \final{conceptual} system setup of an off-the-shelf thin lens and an off-aperture DOE. 
    (Bottom-left) The fitted PSNR plots for recovered images across various DOE locations shows that the optimal location is a trade-off point between the aperture and the sensor. The optimal DOE location for each dataset is approximately $0.24$.
    (Right) Comparison of \ac{PSF} amplitudes at several FoVs when the \ac{DOE} is placed at different locations. The PSFs are cropped by the central 99 pixels for visualization.
    }
    \label{fig:DOE-positions}
    \vspace{-13pt}
\end{figure*}

\subsection{Loss Functions}

During training, the decoder in the network generates multi-scale images used to compute losses tailored to each task. 
For sharp image reconstruction, we employ the pixel-wise \ac{MSE} and the VGG16 perceptual loss $\loss_{\text{percep}}$~\cite{johnson2016perceptual} to ensure the reconstructions closely match the ground truth while maintaining visual fidelity. Given an image patch of size $M \times N$, the \ac{MSE} loss is calculated as $\loss_{\text{mse}} = \frac{1}{M N} \|I - \hat I\|_2^2$.
For depth estimation, we utilize the $\loss_1$ loss which promotes sparsity, defined as $\loss_1 = \frac{1}{M N} \|I - \hat I\|_1$.
To enhance focus power, we define a region $O$ where the majority of the \ac{PSF} energy should be concentrated. This region is a circle with a diameter equal to half the \ac{PSF} size. The \ac{PSF} loss is defined as $\loss_{\text{focus}} = \sum_{(x,y)\not\in O} K(x,y)$.

Notably, multi-task reconstruction requires precise tuning of loss function weights for each task. Instead of relying on heuristic-based tuning methods~\cite{liu2022investigating}, we adopt the \ac{DWA}~\cite{liu2019end} to dynamically adjust the weights \final{$c_w ( w=0,1,\dots)$} based on the average gradients from previous iterations. 
This strategy avoids complex hyperparameter tuning when all losses hold similar importance. Consequently, the total loss function is a weighted sum of individual losses, as follows:
\begin{equation}
    \loss_{\text{total}} = c_{0} \loss_{\text{mse}} + c_{1} \loss_{\text{percep}} + c_{2} \loss_{1} + c_{3} \loss_{\text{focus}} .
\end{equation}

\section{Wide-FoV Simple Lens Imaging}
\label{sec:app1}

\subsection{Design Overview}
Wide-FoV imaging is increasingly challenging when aiming for a compact lens form factor, such as a single refractive lens element. State-of-the-art coded-aperture and/or deep optics imaging solutions mostly assume spatial-invariant PSF behavior and demonstrate in experiments a FoV up to \qty{30}{\degree}. A detailed summary of specifications in relevant work is presented in the supplementary material Tab.~\supp{S1}. 
To this end, our wide-FoV imaging setup includes a simple focusing lens with an approximate FoV of \qty{45}{\degree} and a rotational-symmetric off-aperture DOE, effectively compensating for most aberrations.
Due to computational constraints, the camera is configured with an \ac{EFL} of \qty{35}{\mm} and an f-number of $12$ for proof of concept. Objects are focused at $\qty{1.4}{\m}$ depth. 

The system is trained on a total of 26,281 images sourced from three datasets, which include 21,818 images from the simulated Sceneflow FlyingThings3D (SF)~\cite{mayer2016large}, alongside 2,506 images from the real-world Dualpixels (DP)~\cite{garg2019learning}, and 1,957 images from the Instereo2K (IS)~\cite{bao2020instereo2k}.
During testing, we utilize 1,000 images from the SF dataset, 684 images from the DP dataset, and 50 images from the IS dataset, all upsampled and cropped to match the sensor resolution.

For this scenario, the network employs a single image-head with $c_2=0$. The learning rates are initialized at $1 \times 10^{-4}$ and $6 \times 10^{-3}$ for the network and DOE, respectively.
Training is conducted over 50 epochs, incorporating a cosine annealing schedule starting at epoch 35, reducing the learning rate to $1/10$ of its initial value. Following this, the system undergoes fine-tuning at the adjusted learning rate for an additional 10 epochs, with early stopping triggered after convergence.
We randomly sample two angles for each forward iteration in the off-aperture simulations, aiding in gradient computation within the overlapping regions between each bundle of light. The training process typically takes a day on an Nvidia RTX 4090 GPU.

\begin{figure}[tb]
\centering
\includegraphics[width=\linewidth]{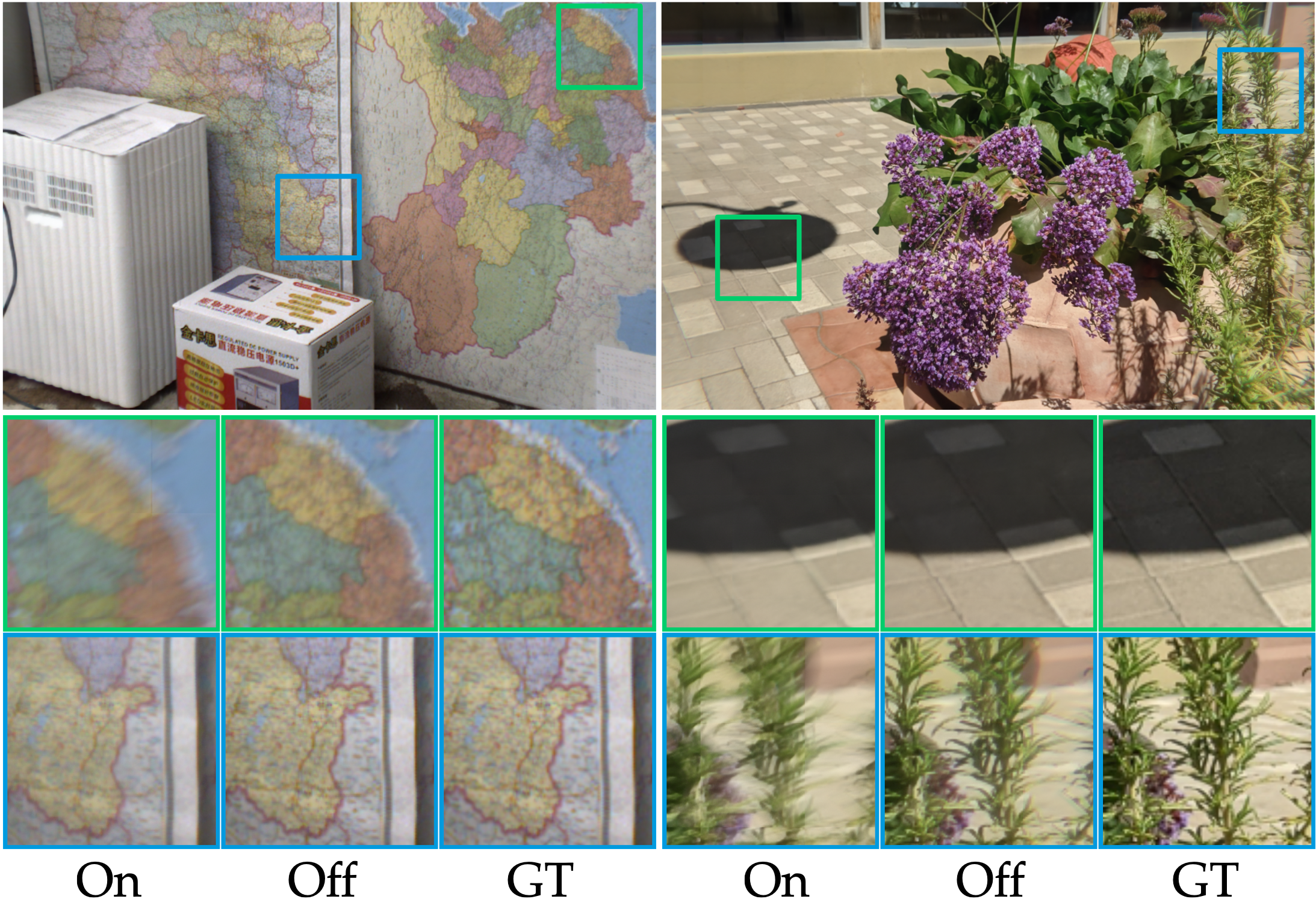}
\vspace{-12pt}
\caption{Recovered 4K-resolution real-world scenes with wide-FoV imaging in simulation. The off-aperture DOE leads to much clearer results than the on-aperture DOE in both central and periphery FoV regions.}
\label{fig:app1-visual}
\vspace{-12pt}
\end{figure}

\subsection{Analysis of DOE Position}

To assess the trade-off between the angular de-multiplexing and local degree of freedom, we conduct an analysis on the system's performance with various DOE positions, as in Fig.~\ref{fig:DOE-positions}.
The DOE is optimized at uniform intervals from 0 to 1 along the optical path, where $0$ corresponds to the principal plane where light is refracted and $1$ to the sensor plane.
Considering assembly difficulties and to mitigate potential performance degradation associated with placing the DOE too close to the sensor, where local degree of freedom is compromised, our analysis focuses on DOE positions of $0\!\sim\!0.4$, corresponding to distances of $\qty{35}\sim\qty{14}{\mm}$ from sensor, respectively.
A spline curve is fitted to the sampled data points using a smoothing factor of $1.7$~\cite{dierckx1995curve}.
\final{A scaled illustration of the system is presented in the Supplementary Material.}


The optimized \ac{PSF} presented on the right of Fig.~\ref{fig:DOE-positions} imply that the angular compromise diminishes as the \ac{DOE} is positioned further away from the aperture (top row).
For a larger \ac{FoV} of $40^\circ$, the \ac{PSF} is the most concentrated around positions $0.2$ and $0.3$. This observation is reinforced by the quantitative results depicted in the plot, indicating the image \ac{PSNR} of DOEs optimized at various positions when assessed across all  datasets.
The off-aperture DOE system achieves \ac{PSNR} values of $35.49~\dB$, $35.44~\dB$, and $31.74~\dB$ on the SF, IS, and DP datasets, respectively. 
This outperforms the on-aperture system by $5.35~\dB$ (SF), $4.41~\dB$ (IS), and $4.13~\dB$ (DP), respectively.
These results are detailed in Tab.~\supp{S2 and S3} in the supplementary material for a comprehensive presentation.
The observed trend underscores the trade-off between local and global degree of freedom, implying that the optimal DOE position should lie between the aperture and sensor planes, but not at either extreme.

Notably, while the \ac{DOE} position could be co-optimized as a parameter, the above analysis suggests that the performance impact is minimal when the DOE is positioned between $0.1$ and $0.3$ of the \ac{EFL}.
As a result, the \ac{DOE} positions are not stable over multiple optimization runs, but all found solutions have comparable overall performance levels. 
For our physical prototypes we therefore opted to abandon the automatic optimization of DOE position and instead chose a position within the high-performance region that is also convenient for mechanical mounting.

\subsection{Simulation Results}

We partition the entire FoV into smaller local patches where shift invariance can be reasonably assumed. Each patch consists of $256 \times 256$ pixels, and we stitch together $16 \times 10$ patches to generate the final 4K resolution images. To mitigate stitching artifacts, we incorporate a $32$-pixel overlap at each boundary.
Figure~\ref{fig:app1-visual} compares the recovered images based on on-aperture and off-aperture DOE using the Instereo2K dataset. These results demonstrate that the off-aperture DOE significantly enhances visual quality in both the central and peripheral regions of the FoV, consistent with the corresponding metrics and optimized PSF characteristics.

\final{While DOEs still encounter challenges in correcting off-axis aberrations, our preliminary results demonstrate their potential over an extended field of view. Specifically, at a \qty{60}{\degree} FoV, the proposed design achieves a $2.5\dB$ improvement compared to on-aperture configurations (see Sec.~S6).}

\section{Wide-FoV Compound Lens RGBD Imaging}
\subsection{Design Overview}

\begin{figure*}[tb]
\centering
\includegraphics[width=0.99\linewidth]{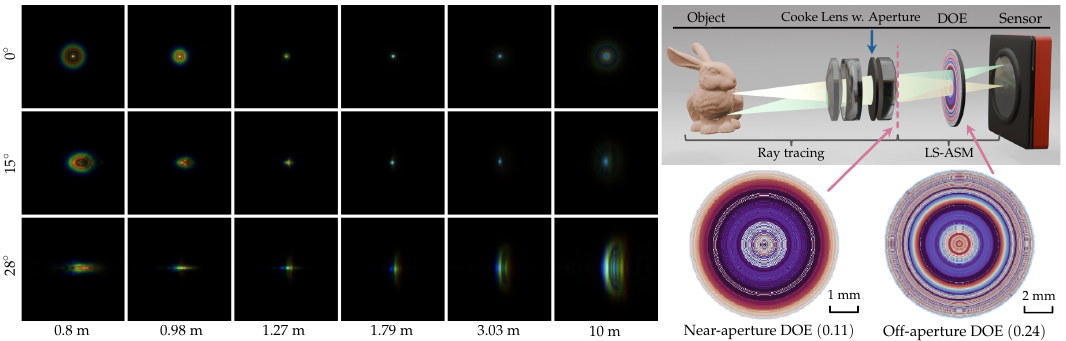}
\vspace{-3pt}
\caption{(Left) Optimized PSFs at depths from \qty{0.8}{\m} to \qty{10}{\m} and FoVs up to \qty{28}{\degree}, using the off-aperture large-FoV EDoF imaging system. Depth layers are sampled uniformly. (Top-right) The 3D model of the optimized system, where the dotted line denotes the location to place the near-aperture DOE. \final{Figure visualization is not drawn to scale.} (Bottom-right) Optimized DOE height maps for near (around $0.11$) and off aperture (around $0.24$) settings. 
}
\vspace{-6pt}
\label{fig:Cooke-psf}
\end{figure*}

While the base application is demonstrated using a simple thin lens, a majority of imaging systems employ compound optics.
These systems, though advantageous, introduce several challenges in our study. First, the aperture plane often does not coincide with the principal plane where light refraction occurs. Second, simulating a \ac{DOE} prior to ray-traced lenses complicates the optimization process. Third, the joint optimization of refractive and diffractive lenses has received limited attention in prior research.

To address these challenges, we introduce a compound-optics prototype designed for wide-FoV depth and color imaging. The system combines an optimized Cooke triplet as the focusing lens module, achieving a practical balance between compactness and aberration correction.
The lens profiles are initially optimized in Zemax OpticStudio to minimize aberrations across the entire \ac{FoV} at the focal distance. 
For ease of fabrication, the last two lenses are replaced with the closest commercially available options, while the first lens is fine-tuned. Additionally, the first surface of the first lens is designed with an aspherical shape to better control off-axis light rays. Detailed specifications of our designed lenses are presented in the supplementary material Tab.~S5. The off-aperture DOE is applied after the lens module.


The camera is configured with a \qty{35}{\mm} \ac{EFL} and an f-number of $7$. To integrate the lens module effectively, we optimize the DOE's position within the back focal length, placing it \qty{22}{\mm} from the sensor but before the pre-assembled sensor components for ease of simulation and assembly. The system is designed to capture both depth and color information for objects within a range of \qty{0.8}{\m} to \qty{10}{\m}, corresponding to approximately $1.2$ Diopter, with a maximum FoV of around $30$ degrees.

\new{We still utilize the three datasets comprising image-depth pairs with diverse features and varying depth map qualities.} For this multi-task application, the network architecture employs two decoding heads: one for \ac{AiF} imaging and another for depth estimation. Based on the optimized compound lens module for \ac{AiF} imaging and insights on balancing multi-tasks from preliminary studies~\cite{liu2022investigating}, we opt to yield a higher loss weight for the more challenging depth estimation task. Accordingly, we assign additional weights on the basis of \ac{DWA} $c_2 = 1$ for depth estimation, while setting $c_0=c_3=0.05$ for other parts. To manage GPU memory constraints associated with larger apertures, we disable perceptual loss by setting $c_1=0$. Experimental results demonstrate a relatively balanced performance between the two tasks.
The training process retains the setting from previous simulations but with the cosine annealing scheduler starting at epoch 15 to gradually reduce the learning rate to $1/100$ of its initial value, ensuring training stability. It typically takes two days to train on the same GPU as before for the increased sampling.

\subsection{PSF Generation}

While the off-axis wave propagator \ac{LS-ASM} has proven effective in generating \ac{PSF}s for simple lenses, it is inadequate for simulating the \ac{PSF}s for compound optics systems. Similarly, \dO~can simulate geometric propagation for compound systems but fails to model fine details from light diffraction. The fusion of these two parts \new{was recently demonstrated in concurrent work~\cite{yang2024end}, while our additional design space requires considering DOE placement}, and the absence of ground truth data for such systems hinders insightful comparisons. In supplementary Fig.~\supp{S3}, we compare the \dO~\ac{PSF}s with those generated from the fused system, showcasing consistent shapes and structures across the entire \ac{FoV}.

\begin{figure}[tb]
\vspace{-7pt}
\includegraphics[width=0.995\linewidth]{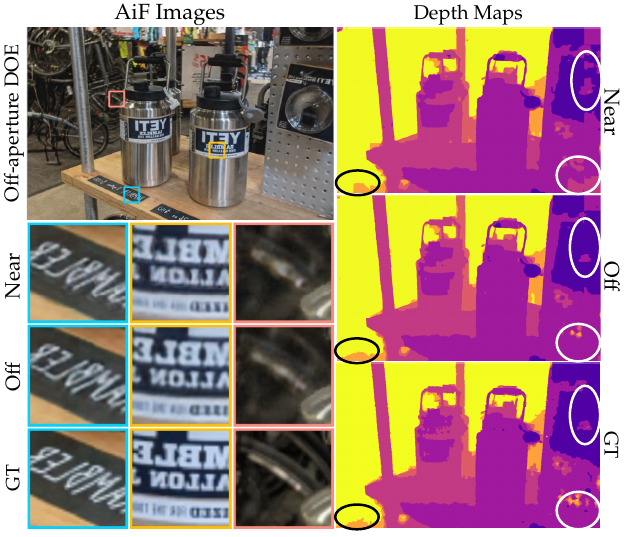}
\vspace{-15pt}
\caption{Recovered 4K-resolution real-world image and depth map with large-FoV EDoF imaging. Zoom-in patches of AiF images of near- and off-aperture DOE systems are presented and differences observed among the obtained depth maps are highlighted.}
\vspace{-13pt}
\label{fig:Cooke-visual}
\end{figure}

\subsection{Simulation Results}
\label{sec:simulation results}

\begin{table}[tb]
\centering
\footnotesize
\renewcommand\arraystretch{1.3}
\tabcolsep=0.2cm
\caption{Assessment of color images in \ac{PSNR} (dB)$\uparrow$ and depth maps in MAE$\downarrow$ when optimizing DOEs at two positions closer and farther from the aperture in the application of wide-FoV depth and color imaging, and when using a simple network. These assessments are conducted on three datasets. The percentages of reduction in MAE of depth maps are reported.}
\label{tab:app2-quant}
\vspace{-6pt}
\addtolength{\tabcolsep}{-3pt}
\begin{tabular}{l|c|c|c}
\toprule
 Dataset & \makecell{Near-aperture~\new{\shortcite{zhuge2024calibration}} \\ + Multi-head} & \makecell{Off-aperture \\ + U-Net~\shortcite{liu2022investigating}} & \makecell{Off aperture \\ + Multi-head} \\
\midrule
Sceneflow & 31.00 / 0.037 & 25.27 / 0.046 & 32.09 / 0.033 (\textbf{-10.8\%})\\
Dualpixel & 27.95 / 0.025 & 23.69 / 0.027 & 28.28 / 0.019 (\textbf{-24.0\%})\\
Instereo2K & 30.51 / 0.032 & 26.06 / 0.124 & 31.42 / 0.027 (\textbf{-15.6\%})\\
\bottomrule
\end{tabular}
\vspace{-10pt}
\end{table}

To conserve computational memory and reduce runtime, we opt not to train the refractive lens with the DOE E2E, as double wave propagation at large angles requires a very high sampling rate. 
Although E2E optimization of a fused system is theoretically feasible, as demonstrated in concurrent work~\cite{yang2024end}, it is orthogonal to the scope of this work. 
Instead, we present a proof-of-concept system that demonstrates the effectiveness of our approach while keeping complexity to a minimum. 

We first design a Cooke triplet using Zemax OpticStudio, which minimizes aberrations at discrete angles (\qty{0}{\degree}, \qty{5}{\degree}, \qty{10}{\degree}, \qty{14}{\degree}) for a focal depth of \qty{1.7}{\m}. 
This optimization yields a lens module with relatively low aberrations across the entire \ac{FoV}, albeit not throughout all depths. 
Using this optimized lens as a base, we integrate a DOE primarily to encode depth information.
Based on the findings discussed in Sec.~\ref{sec:app1}, we position the DOE at approximately $0.24$ of the \ac{EFL}, or \qty{8.4}{\mm} from the aperture plane.
Due to the challenges in simulating the placement of the DOE before the refractive optics, we compare this off-aperture configuration with a near-aperture setup, where the DOE is positioned at the tangential surface of the last lens, approximately \qty{4}{\mm} from the aperture plane ($0.11$ of the \ac{EFL}).
Figure~\ref{fig:Cooke-psf} presents the optimized PSFs of the off-aperture system, sampled uniformly at six depths and three angular directions.
Notably, the system is trained on seven discrete angles (\qty{0}{\degree}, \qty{2.5}{\degree}, \qty{5}{\degree}, \qty{7.5}{\degree}, \qty{10}{\degree}, \qty{12}{\degree}, and \qty{14}{\degree}),
with intermediate angles \qty{1.25}{\degree}, \qty{3.75}{\degree}, \qty{6.25}{\degree}, \qty{8.75}{\degree}, \qty{11}{\degree}, and \qty{13}{\degree} included during testing.
The optimized DOEs for near- and off-aperture systems exhibit distinct characteristics in their outer peripheral rings, highlighting the off-aperture DOE's ability to efficiently encode high-frequency information for large-FoV rays without interference from the inner FoV. This represents a key advantage of the off-aperture design.

Table~\ref{tab:app2-quant} presents a quantitative comparison of the near- or off-aperture systems. 
The off-aperture configuration achieves approximately \qty{1}{\dB} higher \ac{PSNR} for color images and reduces \ac{MAE} for depth maps by up to $24\%$ across three representative datasets. 
The greater improvement in depth maps can be attributed to two factors: First, the near-aperture system's DOE placement at $0.11$ of the \ac{EFL} incurs fewer disadvantages compared to the $0.24$ location, and second, the Cooke triplet is optimized primarily for image performance, leaving more room for improvement in depth recovery.
We further compare the multi-head network with a simple U-Net-based architecture~\cite{liu2022investigating}, which outputs image and depth from the same layer as multiple channels. Our model outperforms the U-Net architecture in both image and depth recovery.

Figure~\ref{fig:Cooke-visual} visually compares the near- and off-aperture systems at 4K resolution. The off-aperture system demonstrates notably superior sharpness in recovered color images compared to its near-aperture counterparts. This enhanced image quality is evident across various depths and throughout FoV, including both inner and outer regions. 
Moreover, the depth map produced by the off-aperture system exhibits significantly enhanced fidelity, particularly at wider FoVs.
\final{We further measure the performance at several sampled points in the Supplementary material.}
These results align with the prior observation that a distance between $0.2$ and $0.3$ represents an optimal choice for off-aperture designs.

\section{Experiments and Results}
\subsection{Prototyping Details}

\begin{figure}
\includegraphics[width=\linewidth]{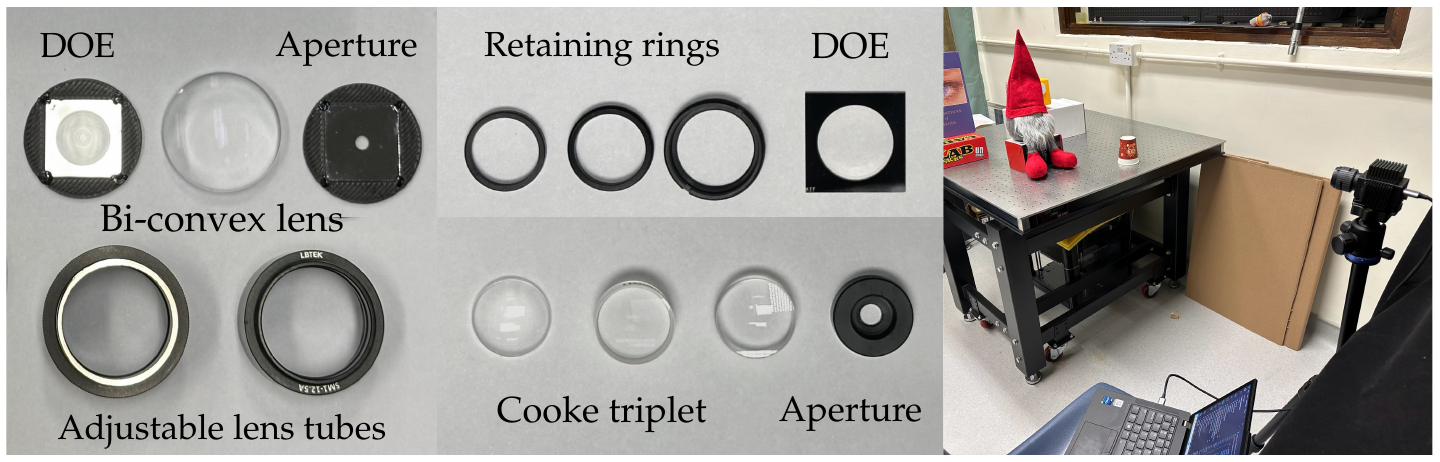}
\vspace{-12pt}
\caption{(Left) The thin lens, DOE, and aperture in App.~1 and (Center) the three refractive lenses, DOE, and other components in App.~2. (Right) The experimental setup.}
\vspace{-9pt}
\label{fig:physical-setup}
\end{figure}

In Application 1, we utilized an off-the-shelf anti-reflex (AR) coated N-BK7 bi-convex lens with a focal length of \qty{35}{\mm} (Thorlabs LB1811-A). Given the wide FoV, we employed a Sony A6400 mirrorless camera equipped with an APS-C sensor. The DOE has an effective diameter of approximately \qty{12}{\mm} and a full diameter of \qty{16}{\mm}. It is mounted with a 3D-printed circular mount with a diameter of 1~inch, in conjunction with the bi-convex lens. The distance between the center plane of the lens and the DOE plane is set at \qty{10.5}{\mm}. The total length of the imaging system outside the camera body is only \qty{21.5}{\mm}, resulting in a compact and lightweight wide-FoV imaging system. A mount adapter has been designed and 3D-printed to connect the SM1 lens tubes to the camera body.

In Application 2, we utilized one AR-coated N-SF11 bi-concave lens of \qty{-15}{\mm} focal length (LD2060-A) and one CaF2 Positive Meniscus Lens of \qty{50}{\mm} focal length (LE5243), and importantly, fabricated one aspherical lens employing \ac{CNC} machining capable of 5-axis single point diamond turning, akin to the techniques reported in state-of-the-art works~\cite{tseng2021differentiable,peng2019learned}. Given the available turning tool, we opt for \ac{PMMA} as the substrate material, implying a refractive index of 1.492 at the principal wavelength of \qty{550}{\nm}. We employ the FLIR GS3-U3-123S6C-C sensor with 4K resolution.

Our DOE fabrication employs an iterative photolithography and dry etching approach to create $2^4$ discrete phase levels on a fused silica wafer, with a Chromium layer acting as an optical baffle~\cite{fu2021etch,dun2020learned}. Repeating the photolithography and reactive-ion etching steps four times preserves the high-frequency spatial features crucial for the DOE design. Detailed fabrication parameters are provided in the supplementary material.
We then assembled all optics with customized focal-tuning mounts. 
The setup diagrams and photographs of the components in both applications are shown in Fig.~\ref{fig:teaser} (center) and Fig.~\ref{fig:physical-setup}.

\begin{figure*}[ht]
\includegraphics[width=\linewidth]{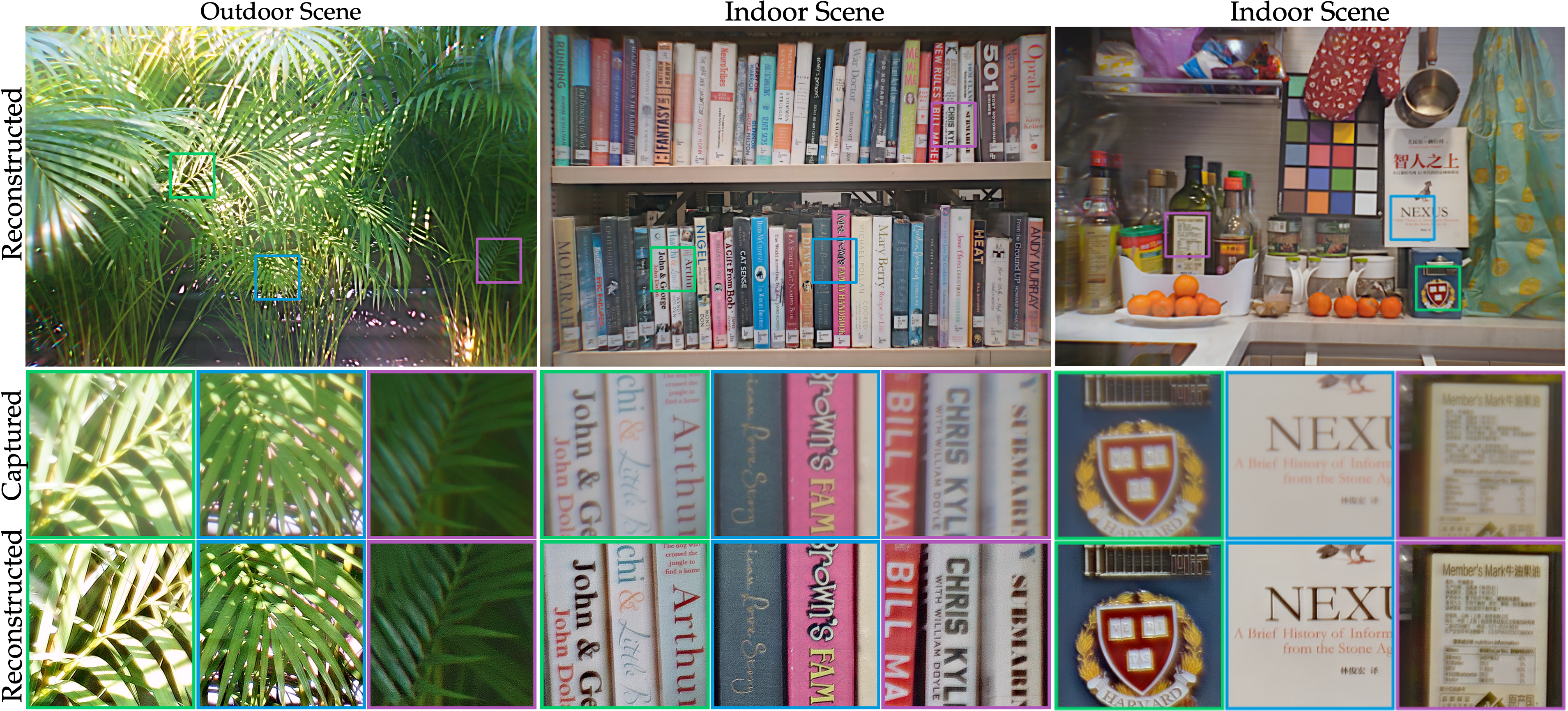}
\vspace{-12pt}
\caption{Indoor and outdoor experimental results for the application of wide-FoV simple lens imaging. 
The equivalent f-number is 12. ISO is set as 100 for all captures. 
}
\vspace{-6pt}
\label{fig:real-capture-app1}
\end{figure*}

\subsection{PSF Calibration and Network Fine-tuning}

We employ a white laser source (LS-WL1 from Edmund Optics) coupled with collimating lenses and a $\qty{25}{\um}$ pinhole to serve as a point light source. 
In the first application, the PSFs captured on the sensor at a distance of \qty{1.4}{\m} are measured at 24 distinct locations within a quadrant of the image plane. Each measured PSF is spaced approximately 512 sensor pixels apart. 
In the second application, we capture \ac{PSF}s at five \ac{FoV}s for each depth plane of the nine depths from $\qty{0.8}{\m}$ to $\qty{5}{\m}$. A detailed comparison between the simulated and captured PSFs is provided in the Supplementary Material.

Upon capturing the \ac{PSF}s, we simulate measurements and fine-tune the network using the three datasets. 
However, we observe strong haze effects in the measurements of App.~2, which are likely caused by imperfections in the fabrication and assembling process~\cite{li2023extended}. 
To address this issue and enable the network to learn optical encodings from the corrupted measurements, we manually perturb the captured PSFs by introducing a zero-frequency term and applying random scaling to simulate realistic measurement conditions. 
Furthermore, we enhance the tuning process by incorporating a high-performing semantic-based depth estimation model ``Depth Anything''~\cite{yang2024depth}. 
Despite this, our optical encoded system successfully predicts plane depth maps from fake scenes displayed on a monitor, whereas semantic models fail to do so, highlighting the continued importance of optical encoding in resolving RGBD imaging. Further details and visualization are available in the supplementary material Fig.~\supp{S13}.


\subsection{Real-world Results}

\begin{figure*}
\includegraphics[width=\linewidth]{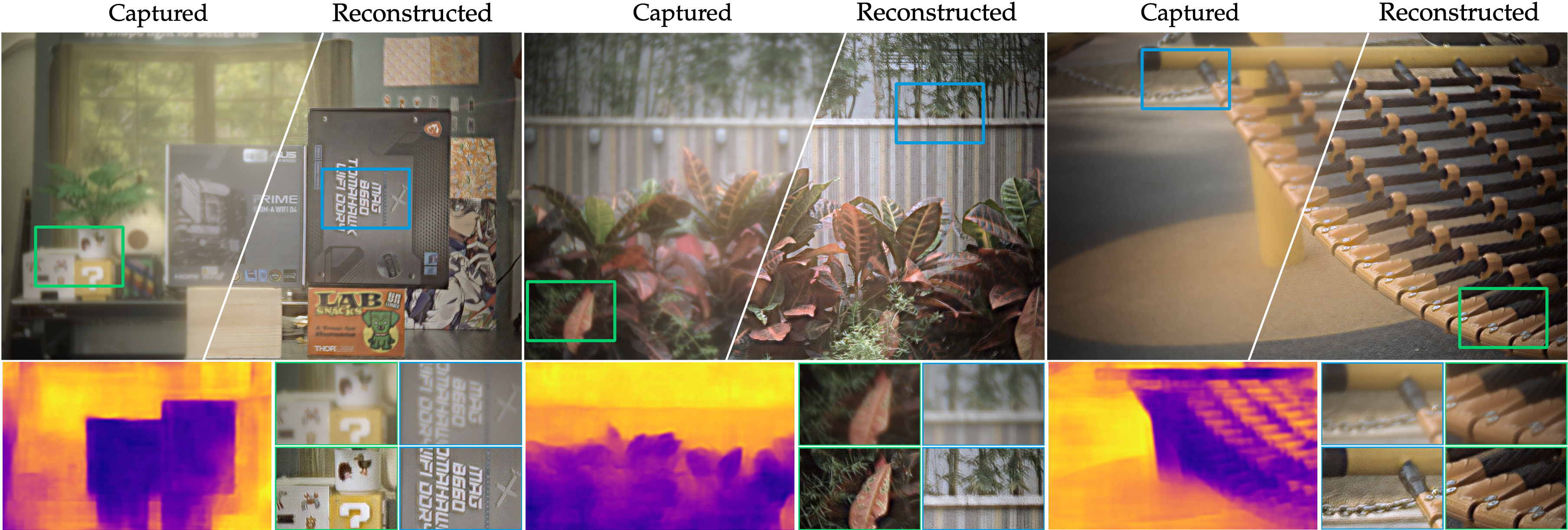}
\vspace{-12pt}
\caption{Indoor and outdoor experimental results for App.~2: wide-FoV \ac{AiF} and depth imaging. For each scene, we present the captured and reconstructed halves of color images (top), along with their zoom-in patches (bottom-right) and estimated depth map (bottom-left). The equivalent f-number is 7, and the gain set 0 for all captures. }
\vspace{-9 pt}
\label{fig:real-capture-app2}
\end{figure*}

The experimental results from real-world indoor and outdoor scenes, captured using the customized camera prototypes, are presented in Fig.~\ref{fig:real-capture-app1} for App.~1 and Fig.~\ref{fig:real-capture-app2} for App.~2, as well as in Fig.~\ref{fig:teaser}. 

For App.~1, the reconstructed images exhibit significant improvements over the raw measurements across both inner and outer FoVs. 
High-frequency details, such as text and edges, are notably sharper in the reconstructed images. For instance, even small characters, like those on bottle labels, are accurately recovered, underscoring the method's capability to reconstruct wide-FoV information effectively.

In App.~2, the \ac{AiF} images demonstrate sharpness across the entire depth range. Notably, distant objects, such as highly blurred tree leaves that are otherwise unrecognizable to human eye, are faithfully reconstructed. Additionally, the recovered depth maps align well with visual cues, displaying clear transitions between depth planes. For example, the scene on the left shows distinct edges between the box and its background, while the right scene highlights depth variations in the net's hollows. These results validate the effectiveness of our approach in wide-FoV AiF imaging and depth encoding.
Additional real-world results of both camera prototypes are presented in Supplementary Fig.~\supp{S7}.



\new{
\noindent
\textbf{Comparisons with Simple Lens and Commercial Compound Lens.}
Although prior works have extensively demonstrated the advantages of DOE-based imaging systems, we also validate our App.~1 prototype against two baselines: (1) an off-the-shelf achromatic doublet with a pre-trained restoration network~\cite{chen2022simple} and (2) a commercial Sony digital single-lens reflex (DSLR) lens. 
Compared to the first baseline, our DOE-encoded system shows superior detail preservation, particularly in wide-FoV regions.
Against the compound lens, our off-aperture design achieves comparable performance in diverse indoor and outdoor scenes, despite its significantly smaller form factor. 
\final{A simulated non-coded thin lens system is also compared by metrics.}
Additional details are provided in Supplementary Sec.~\supp{6}.
}

\subsection{Bells \& Whistles: Comparison with A Pinhole}

\begin{figure}[tb]
    \centering
    \includegraphics[width=\linewidth]{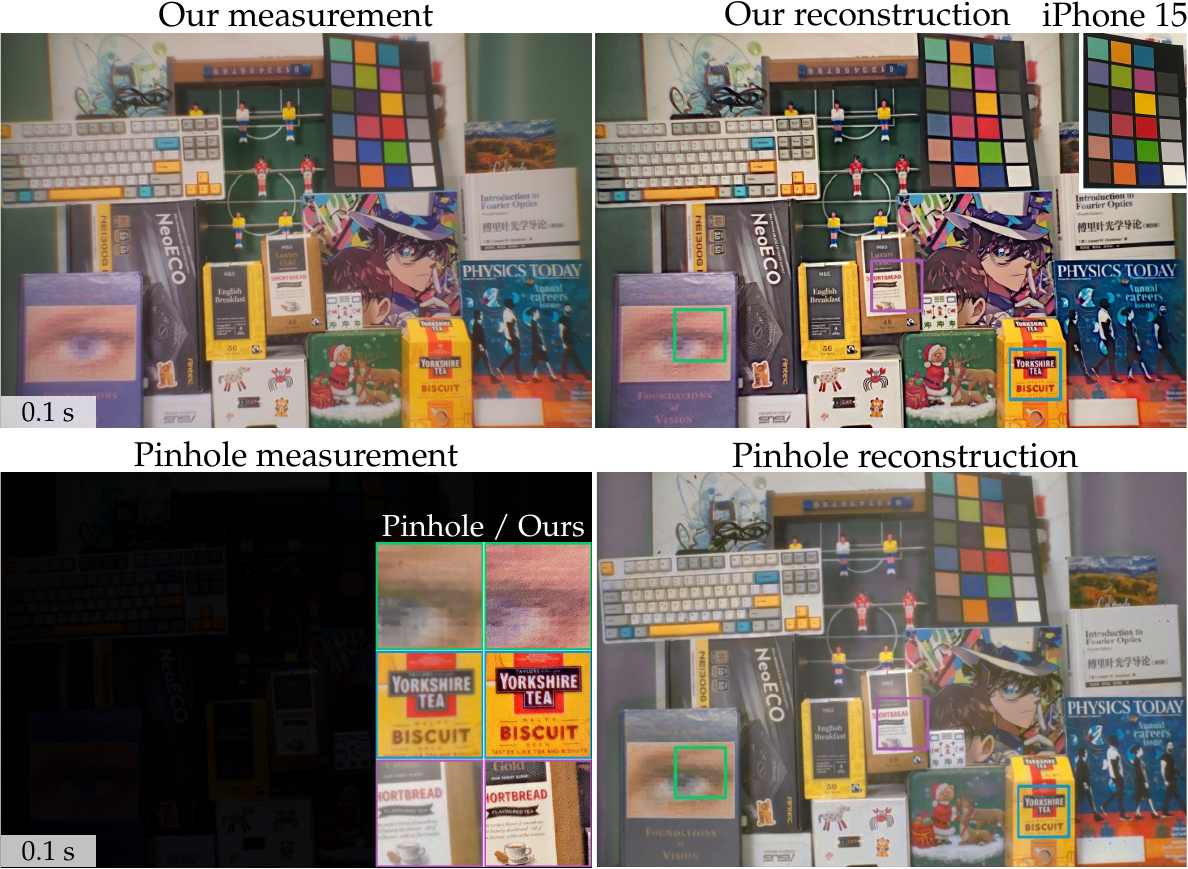}
    \vspace{-9pt}
    \caption{Experimental results of our wide-FoV RGB imaging prototype compared with denoised~\cite{chen2018learning} results from a pinhole camera setting (0.5~mm diameter). Ours recovers correct color tones (with a reference captured by iPhone 15) and sharper edges. Both are captured with a shutter speed of \qty{0.1}{\sec} and an ISO of 500 under indoor illumination.}
    \label{fig:pinhole}
    \vspace{-12pt}
\end{figure}

We further evaluate our system's wide-FoV imaging performance in App.~1 by comparing it with a theoretically aberration-free pinhole setup. 
In this experiment, an indoor scene was captured under fixed exposure conditions, illustrated in Fig.~\ref{fig:pinhole}. 
The counterpart pinhole setup employs the same convex lens but is paired with a \qty{0.5}{\mm} iris aperture. While this configuration minimizes severe off-axis aberrations at wide angles, it also significantly degrades the signal-to-noise ratio. We observe that the measurement appears extremely dim.
Even after applying a state-of-the-art low-light image restoration algorithm~\cite{chen2018learning} to the pinhole measurement, the result shows inferior detail recovery compared to our method, which maintains sharp edge fidelity across the entire FoV and ensures closer-to-real color reproduction. Importantly, our prototype's measurement preserves a higher fidelity of natural color tone, as evidenced by the color checkerboard captured with a cellphone camera.

Furthermore, while pinholes can reduce off-axis aberrations, large aperture settings remain essential for computational visual tasks that rely on optical encoding~\cite{zhou2011coded}, such as RGBD imaging demonstrated in App.~2. In this context, pinhole configurations inherently extend the depth of field to near infinity, rendering depth recovery based on PSF encodings less efficient. This limitation can restrict the system to a purely semantic-based depth estimation approach, which is orthogonal to the scope of our off-aperture encoding exploration.

\section{Discussion and Conclusion}
In conclusion, we have presented an off-aperture \new{co-optimized} imaging system designed for wide-FoV imaging applications. By positioning the DOE at a plane distinct from the aperture or pupil plane, we have achieved enhanced angle-dependent encoding through an additional design parameter. 
Our analysis of the impact of DOE placement provides valuable insights into how system performance is influenced by the DOE's position \new{and demonstrates a region of jitter-robustness}. 
\new{Supplementary material also includes an optimization curve, which converges to $\sim0.3$ in our experimental attempts, validating our choice.}
By decoupling light from different angles, we can enhance off-axis aberration correction by over 5~dB in PSNR. 

Importantly, we have validated our system with two custom-designed and assembled camera prototypes: one featuring a simple lens and the other fusing multiple refractive optics, both equipped with \new{co-optimized}, off-aperture-positioned DOEs. 
These prototypes have demonstrated the system's effectiveness in RGB-only and RGBD imaging applications, achieving a FoV of up to \qty{45}{\degree} and \qty{28}{\degree}, respectively.
We believe that the proposed off-aperture encoding paradigm holds great potential for broader applications, such as compact and wide-FoV cameras tailored for AR/VR, autonomous systems, and edge devices, where enhanced angular information encoding capability is critical.

\vspace{3pt}
\noindent\textbf{Overview of Limitations:}
While all the individual components of the refractive-diffractive-hybrid, off-aperture encoded imaging framework are differentiable, we present an imaging system that optimizes the refractive lenses and the DOE in two separate steps to accommodate computational constraints, such as limited GPU memory, particularly when facilitating wide-FoV designs. 
In addition, experimental results also reveal that incorporating a \ac{DOE} can introduce artifacts such as halos, which are attributable to limitations in diffraction efficiency and fabrication and assembly tolerances, especially under non-uniform and high illuminance conditions. 
Despite being widely validated on large image datasets, advanced semantic models are more susceptible to visual ambiguities, rendering them a practical yet imperfect choice for depth ground truth in fine-tuning. A more reliable alternative is to utilize active depth-sensing instruments such as a time-of-flight camera as ground truth.

\ifpeerreview \else
\section*{Acknowledgments}
This work was partially supported by the National Science Foundation of China (62322217), the Research Grants Council of Hong Kong (GRF 17208023 \& 17201822), the Innovation and Technology Fund of Hong Kong (ITP/062/24AP), and the Hong Kong STEM Fellowship awarded to W. Heidrich.
The authors thank Mr. Feifan Qu and Ms. Duolan Huang for assisting with the experimental acquisition.
\fi

\bibliographystyle{IEEEtran}
\bibliography{bib}

\end{document}